\documentclass[aps,prl,twocolumn,showpacs,superscriptaddress,preprintnumbers,amsmath,amssymb,longbibliography]{revtex4-1}

\usepackage[utf8]{inputenc}
\usepackage{graphicx}
\usepackage{bm}
\usepackage{ulem}    
\usepackage{xcolor}
\usepackage{gensymb}                  
\usepackage{tikz}
\usepackage[pdftex,colorlinks=true,pdfstartview=FitV,linkcolor=blue,citecolor=blue,urlcolor=blue]{hyperref}

\begin{document}

\title{Atomic Scale Surface Segregation in Copper-Gold Nanoparticles}
\author{Gr\'egoire Breyton}
\affiliation{Universit\'e Paris Cit\'e, CNRS, Laboratoire Mat\'eriaux et Ph\'enom\`enes Quantiques (MPQ), 75013 Paris, France}
\affiliation{Laboratoire d'Etude des Microstructures, ONERA-CNRS, UMR104, Universit\'e Paris-Saclay, BP 72, Ch\^atillon Cedex, 92322, France}
\author{Hakim Amara}
\affiliation{Universit\'e Paris Cit\'e, CNRS, Laboratoire Mat\'eriaux et Ph\'enom\`enes Quantiques (MPQ), 75013 Paris, France}
\affiliation{Laboratoire d'Etude des Microstructures, ONERA-CNRS, UMR104, Universit\'e Paris-Saclay, BP 72, Ch\^atillon Cedex, 92322, France}
\author{Jaysen Nelayah}
\affiliation{Universit\'e Paris Cit\'e, CNRS, Laboratoire Mat\'eriaux et Ph\'enom\`enes Quantiques (MPQ), 75013 Paris, France}
\author{J\'er\^ome Creuze}
\affiliation{ICMMO/ESP2M, Universit\'e Paris-Saclay, UMR 8182, 17 avenue des sciences, 91405 Orsay cedex, France}
\author{Hazar Guesmi}
\affiliation{ICGM ICMMM - Institut Charles Gerhardt Montpellier - Institut de Chimie Mol\'eculaire et des Mat\'eriaux de Montpellier }
\author{Damien Alloyeau}
\affiliation{Universit\'e Paris Cit\'e, CNRS, Laboratoire Mat\'eriaux et Ph\'enom\`enes Quantiques (MPQ), 75013 Paris, France}
\author{Guillaume Wang}
\affiliation{Universit\'e Paris Cit\'e, CNRS, Laboratoire Mat\'eriaux et Ph\'enom\`enes Quantiques (MPQ), 75013 Paris, France}
\author{Christian Ricolleau}
\email{christian.ricolleau@univ-paris.fr}
\affiliation{Universit\'e Paris Cit\'e, CNRS, Laboratoire Mat\'eriaux et Ph\'enom\`enes Quantiques (MPQ), 75013 Paris, France}

%

\begin{abstract}

In this work, we combine electron microscopy measurements of the surface compositions in Cu-Au nanoparticles and atomistic simulations to investigate the effect of gold segregation. While this mechanism has been extensively investigated within Cu-Au in the bulk state, it was never studied at the atomic level in nanoparticles. By using energy dispersive X-ray analysis across the (100) and (111) facets of nanoparticles, we provide evidence of gold segregation in Cu$_{3}$Au and CuAu$_{3}$ nanoparticles in the 10 nm size range grown by epitaxy on a salt surface with high control of the nanoparticles morphology. To get atomic-scale insights into the segregation properties in Cu-Au nanoparticles on the whole composition range, we perform Monte Carlo calculations employing $N$-body interatomic potentials. These simulations confirm this effect by showing a complete segregation of Au in the (100) and (111) faces of a truncated octahedron for gold nominal composition of the alloy above 70\% and 60\% respectively. Furthermore, we show that there is no size effect on the segregation behaviour since we evidence the same oscillating concentration profile from surface to the nanoparticles core as in the bulk. These results can shed new lights in the interpretation of the enhanced reactivity, selectivity and stability of Cu-Au nanoparticles in various catalytic reactions.

\end{abstract}

\maketitle 

Surface segregation in A$_{x}$B$_{1-x}$ binary alloys, i.e. the enrichment of surface by one of the elements as compared to the bulk composition, has been the subject of numerous studies~\cite{Vasiliev1997, Chelikowsky1984, Balseiro1980, Ponec1976}. It is a very important phenomenon in surface physics of alloys since it can dramatically change the intrinsic properties of the bulk material. Notably, it can strongly modify the surface reactivity of the alloys during catalytic reactions~\cite{Okada2016, Nakanishi1992}. Three physical parameters are used to determine \textit{a priori} the element that segregates at the surface~\cite{Wynblatt1977}. The first two ones are the surface energy and the atomic size of the species that constitute the alloy. It is generally admitted that the element with the lower surface energy and with the larger size will segregate~\cite{McLean1957, Sparnaay1983, Guisbiers2014}. Another driving force is the alloying effect meaning the competition between the cohesive energy of each individual atoms of the alloy and the free energy of mixing. These key quantities explaining segregation phenomena have been put forward by using phenomenological models based on the pair-based~\cite{Williams1974, VanSanten1974, MoranLopez1977} and the elastic-strain energy theories~\cite{McLean1957, Burton1977}. Moreover, numerical calculations were performed within the tight-binding approximation to characterize the surface segregation from a microscopic point a view with an accurate description of the chemical bonds in transition metal based alloys~\cite{Mottet2002, Legrand1990, Treglia1987, Balseiro1980, Kerker1977, Lambin1978}.

In bulk CuAu alloys, this effect has been extensively studied from both experimental and theoretical approaches as a model system for binary alloys~\cite{Sparnaay1983, Guisbiers2014, Mroz2005, Mroz2001, McDavid1975, Buck1983, McRae1990, Reichert1995, Reichert1996, Cheng2006, Wilson2002, Pauwels2001, Ascensio2006}. Among the most common ones, the techniques that were used to probe the surface composition of bulk Cu-Au are Auger electron spectroscopy (AES)~\cite{Mroz2005, Mroz2001, McDavid1975}, low energy ion scattering (LEIS)~\cite{Sparnaay1983, Buck1983, McRae1990}, low-energy electron diffraction (LEED)~\cite{McRae1990}, and X-rays surface diffraction~\cite{Reichert1995, Reichert1996}. All of these works have shown various extent of Au surface segregation depending on the nature of the material (mono- vs polycrystal), the experimental technique and the indexes of the considered surfaces. Interestingly, this segregation effect is followed by an oscillating concentration profile from the surface to the core of the material where the nominal concentration is finally reached~\cite{Buck1983, Reichert1995, Hayoun1998,  Hou1997}.

For nanoparticles (NPs), the situation can be different due to the so-called size effect i.e. the competition between the bulk and surface energies of the NP~\cite{Peng2015} resulting in segregation effects as in case of Ag-Pt~\cite{Pirart2019}, Cu-Ag~\cite{Langlois2012} or Ni-Pt~\cite{Cui2013} NPs. Indeed, physics at the nanoscale could be different than the one occurring in bulk, especially for NPs whose diameter is smaller than around 10 nm. Typical examples include the dependence on the size of the melting temperature~\cite{Buffat1976}, surface energy~\cite{Amara2022} or mechanical properties~\cite{Amodeo2021} of pure NPs as well as the order-disorder transition temperature for bimetallic nanoalloys~\cite{Alloyeau2009}. Regarding segregation effects in Cu$_{x}$Au$_{1-x}$ NPs very few studies exist. Results are mainly obtained by atomistic calculations~\cite{Cheng2006, Wilson2002, Pauwels2001,Ascensio2006, Li2016} and nanothermodynamic approaches~\cite{Guisbiers2014}. All these works demonstrated Au segregation at the surface of the NPs. In one case, it was also demonstrated Cu enrichment of the NPs facets although this configuration is not stable~\cite{Ascensio2006}. From experiments, up to now, two papers report results on this system~\cite{Ascensio2006, Guisbiers2014}. The segregation effect was evidenced by chemical mapping acquired by X-ray spectroscopy using a Transmission Electron Microscopy (TEM) in Scanning mode (STEM). In these works, the Au segregation was revealed at the nanometer scale and in consequence there is no evidence of  the different extent of the segregation on the three main low index facets, namely (111), (110) and (100), as observed in bulk systems. Indeed, such analysis along a NP, which is much more complex, has never been addressed.

In this letter, we determine at the atomic scale the chemical composition of individual facets of epitaxially grown CuAu$_{3}$ and Cu$_{3}$Au NPs on NaCl (100) surface in truncated octahedral shape by using X-ray spectroscopy in an aberration-corrected electron microscope. We then compared the results to Monte Carlo simulations allowing the  determination of the composition of (111) and (100) facets of Cu-Au NPs in the whole composition range. We show a remarkably good agreement between both approaches proving unambiguously the segregation of gold on NPs surfaces at the atomic scale followed by an oscillating concentration profile within the particle.

From an experimental point of view, the challenge to evidence unambiguously the effect of Cu or Au segregation is to have a perfect control on the 3D morphology of the NPs and then on the facets exhibited by the particles under consideration. For that purpose, we developed the epitaxial growth of CuAu NPs on a NaCl substrate and deposited on a TEM carbon grid by the carbon replica technique~\cite{PierronBohnes2014}. Cu-Au NPs were synthesized by alternated pulsed laser deposition technique in a high vacuum chamber under a pressure of 10$^{-8}$ Torr~\cite{Alloyeau2007}. Two compositions were prepared in the in the Cu$_{3}$Au and CuAu$_{3}$ stoichiometry ranges and the growth was made at 400\degree C in order to obtain NPs in the FCC disordered phase (A1 phase). Experimental details are given in Sec. I of the Supplemental Materials. The NPs were imaged by using a double aberration corrected electron microscope (JEOL ARM 200F cold FEG) in STEM mode using the High Angle Annular Dark Field (HAADF) technique. Chemical analysis of the NPs surface composition was performed by Energy Dispersive X-ray spectroscopy (EDX).
\begin{figure}[htbp!]
\begin{center}
\includegraphics[width=1\linewidth]{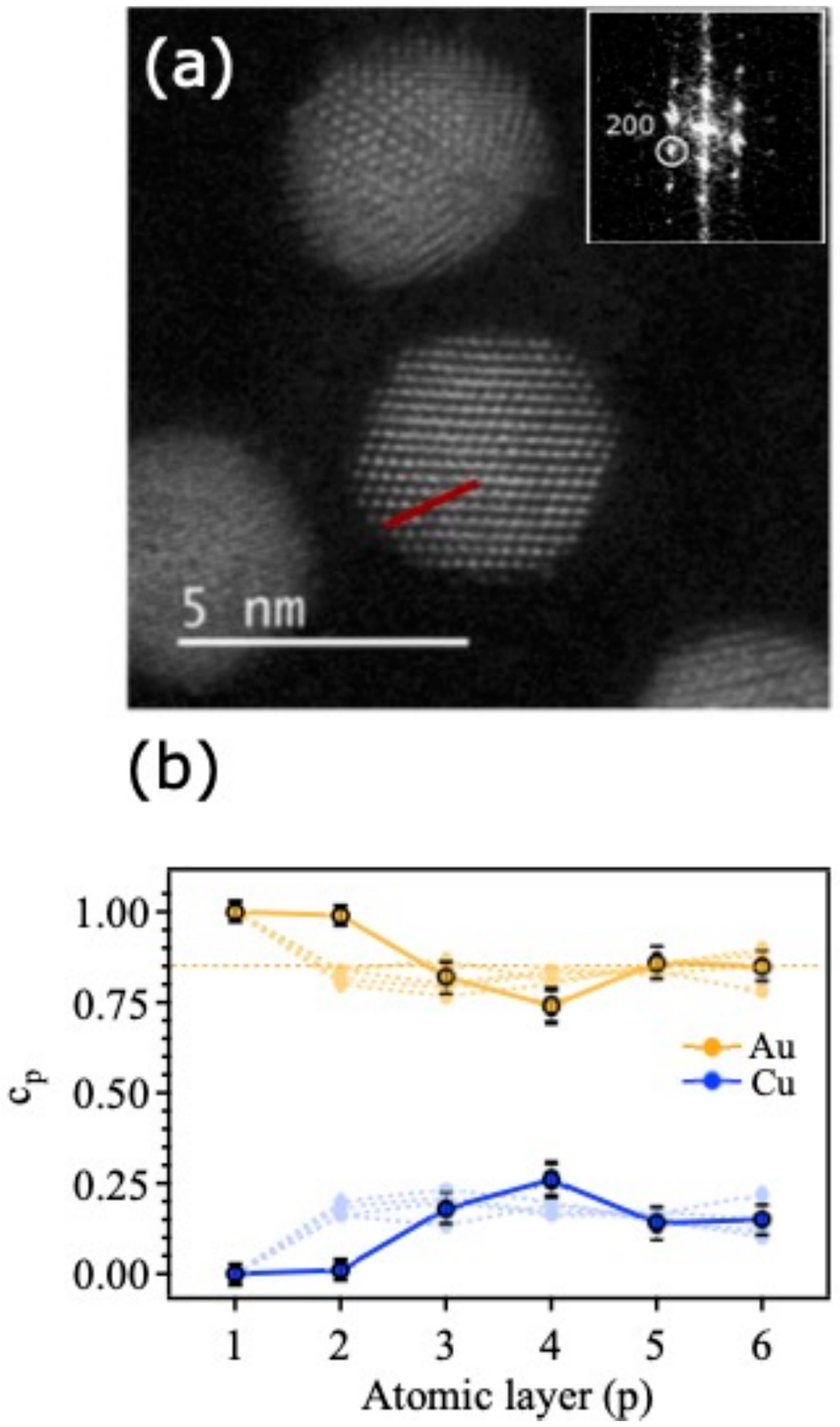}
\caption{(a) HRSTEM image of a Cu$_{15}$Au$_{85}$ NP oriented along the $\left[011\right]$ zone axis. Inset: FFT of the image showing the $\left[200\right]$ direction. Red line: direction of the line scan for the EDX analysis. (b) Composition profile in Au and Cu as a function of the position of the plane along the $\left[200\right]$ direction from the surface (atomic layer 1) to the core of the NP. Light colors: Au and Cu concentrations extracted from the atomistic simulations performed on a NP with the same nominal composition along the same family of planes as the experimental ones.}
\label{Cu15Au85}
\end{center}
\end{figure}

Figure~\ref{Cu15Au85}a shows a typical HAADF high resolution STEM image of a CuAu$_{3}$ stoichiometry NP oriented along the $\left[011\right]$ zone axis. The exact composition measured by EDX spectroscopy over the whole NP is Cu$_{15}$Au$_{85}$. Through this projection, we clearly identify a truncated octahedron exhibiting 6 facets, namely two (200) and four (111) ones, parallel to the electron beam. Since the Fast Fourier Transform (FFT) pattern of the NP does not show any super structure reflections, the NP is in the A1 phase. We analyze the composition of (200) planes from the surface to the core of the NP across a line scan of the beam along the $\left[200\right]$ direction to the planes (red line in Fig.~\ref{Cu15Au85}a). The procedure is described in Sec. II of the Supplementary Material. From each spectrum acquired along this line, we quantified the composition of an atomic column belonging to the (200) planes by analyzing the intensities under the Au-L$_{\alpha}$ and Cu-K$_{\alpha}$ edges using the Cliff-Lorimer method with a theoretical k$_{Cu/Au}$~\cite{Cliff1975} factor. The results are plotted as a function of the position of the beam along the line scan and shown in Fig~\ref{Cu15Au85}b. Since the alloy is in the solid solution phase state, each site is randomly occupied by Cu or Au atoms with a probability of 0.15 and 0.85 respectively. Each column being equivalent in the lack of segregation, its composition is thus equal to the one of the atomic plane. From Fig.~\ref{Cu15Au85}b, it clearly appears that the (200) surface plane (atomic layer 1) is made of pure gold and then the composition tends to Cu$_{15}$Au$_{85}$ for the planes belonging to the core of the NP. It should be pointed out that in this composition range, i.e. in the Au rich region of the phase diagram of the Cu-Au system, when the first surface plane is saturated in gold, the subsequent planes become enriched with copper to reach the nominal composition of the alloy (Fig~\ref{Cu15Au85}b). Note that in addition to the accuracy of the EDX technique without reference sample which is around 5 at. \%, the precision of the analysis is very sensitive to the exact alignment of the plane with respect to the electron beam: a small misalignment may cause that the spectrum does not strictly correspond to the composition of a unique atomic column.

\begin{figure}[htbp!]
\includegraphics[width=1\linewidth]{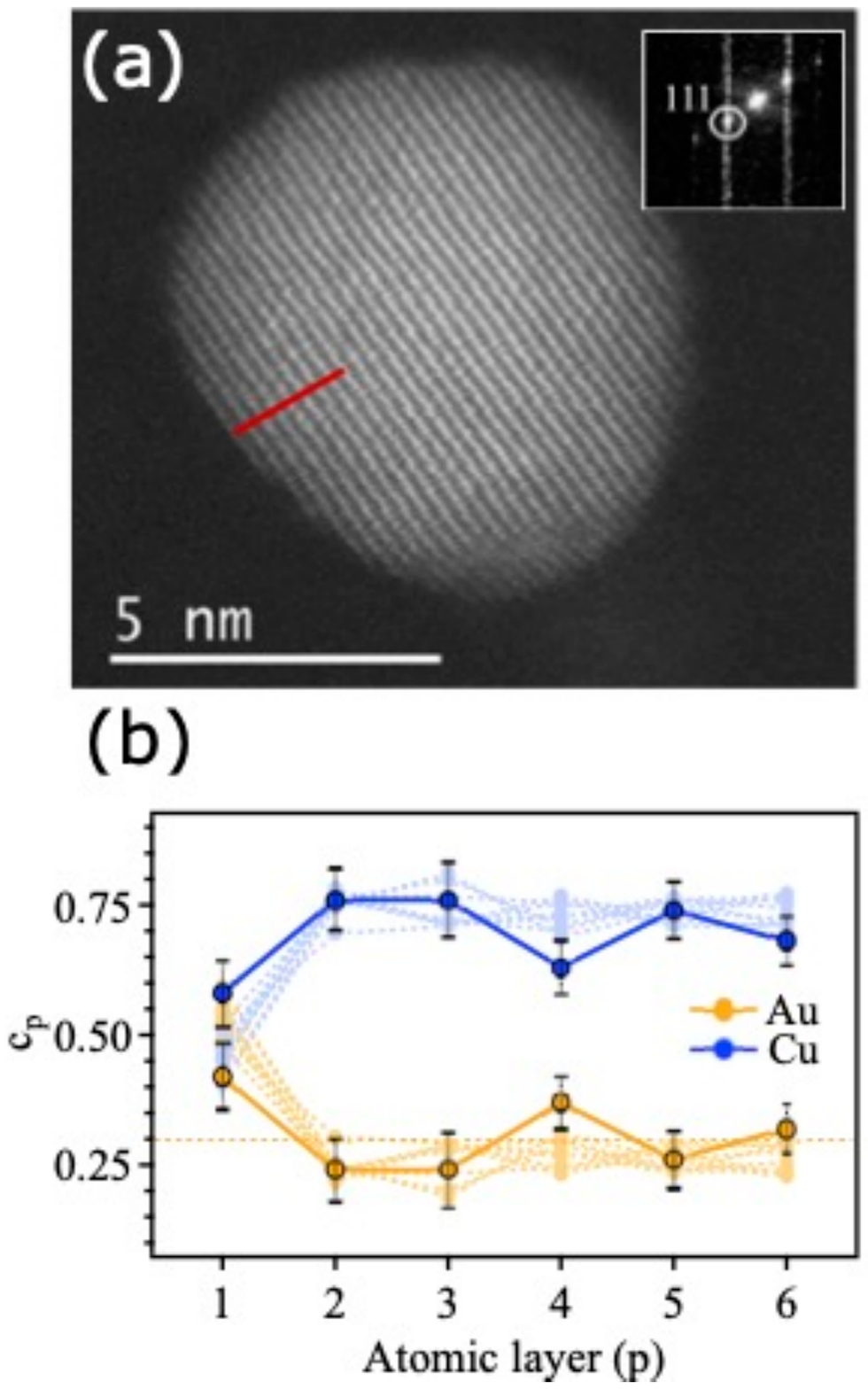}
\caption{(a) HRSTEM image of a Cu$_{70}$Au$_{30}$ NP oriented near the $\left[110\right]$ zone axis. Inset: FFT of the image showing the $\left[111\right]$ direction. Red line: directions of the line scan for the EDX analysis. (b) Composition profile in Au and Cu as a function of the position of the plane along the $\left[111\right]$ direction from the surface (atomic layer 1) to the core of the NP. Light colors: Au and Cu concentrations extracted from the atomistic simulations performed on a NP with the same nominal composition along the same family of plane as the experimental ones}
\label{Cu70Au30}
\end{figure}

For the Cu$_{3}$Au stoichiometry, an HAADF image of a Cu$_{70}$Au$_{30}$ NP is shown in Figure~\ref{Cu70Au30}a. The nanoparticle is oriented near the $\left[110\right]$  zone axis and the corresponding FFT pattern exhibits the 111 reflections. No super structure reflexions are observed confirming that the NP is in the disordered FCC state. Hence, the expected composition of these planes must be Cu$_{70}$Au$_{30}$. We analyzed the plane compositions from the surface to the core along a line scan perpendicular to the (111) surface of the particle following the same procedure as before. The results are plotted in Figure~\ref{Cu70Au30}b. According to these concentration profiles, it appears clearly that the composition of the (111) surface is enriched in gold meaning 42\% instead of 30\% according to the NPs composition and a depletion in copper, 58\% instead of 70\%. The nominal composition of the plane is recovered from the second plane.

To get insight the segregation properties of Cu-Au NPs at atomic scale, we perform Monte Carlo (MC) simulations using a specific $N$-body potential derived from the second moment approximation (SMA) of the tight-binding (TB) scheme~\cite{ducastelle1970, rosato1989}. The interatomic potential is included in a MC code in the canonical ensemble to relax the structures at finite temperatures~\cite{Frenkel2002}. Here, the simulations are performed at high enough temperatures to ensure that the NPs are in a disordered state as the experiments. Regarding the procedure for adjusting the TB-SMA potential and the MC calculations, more details can be found in Sec. III of the Supplemental Material. Meanwhile, we note that the calculated Au surface energies are lower than those of Cu, in line with \textit{ab initio} calculations, favoring Au surface segregation. Since we focus on segregation phenomena, we have ensured that our TB potential can satisfactorily reproduce the enthalpy of segregation of the solute at the surface, $\Delta H^{seg}$. Indeed, the tendency of a constituent to segregate at the surface is characterized by this crucial quantity defined as the energy balance involved when one atom of a given species, initially placed in the bulk, is exchanged with an atom of the other species located at the surface~\cite{Treglia1999, Creuze2015}. A negative value of $\Delta H^{seg}$ means that solute segregation is favored. As shown in Sec. III of the Supplemental Material, our TB-SMA results are in agreement with the \textit{ab initio} data. More precisely, we notice a  strong tendency for the Au impurity to segregate on the first layer and this whatever the Cu surface considered, namely (100), (111) and (101). The conclusions are rather different in the case of the copper solute where an opposite behavior is observed. Even if the alloying effect should obviously not be neglected, we can expect to observe a strong segregation of Au within NPs through our simulations and quantify more accurately this phenomenon at the atomic scale.

\begin{figure}[htbp!]
\includegraphics[width=0.95\linewidth]{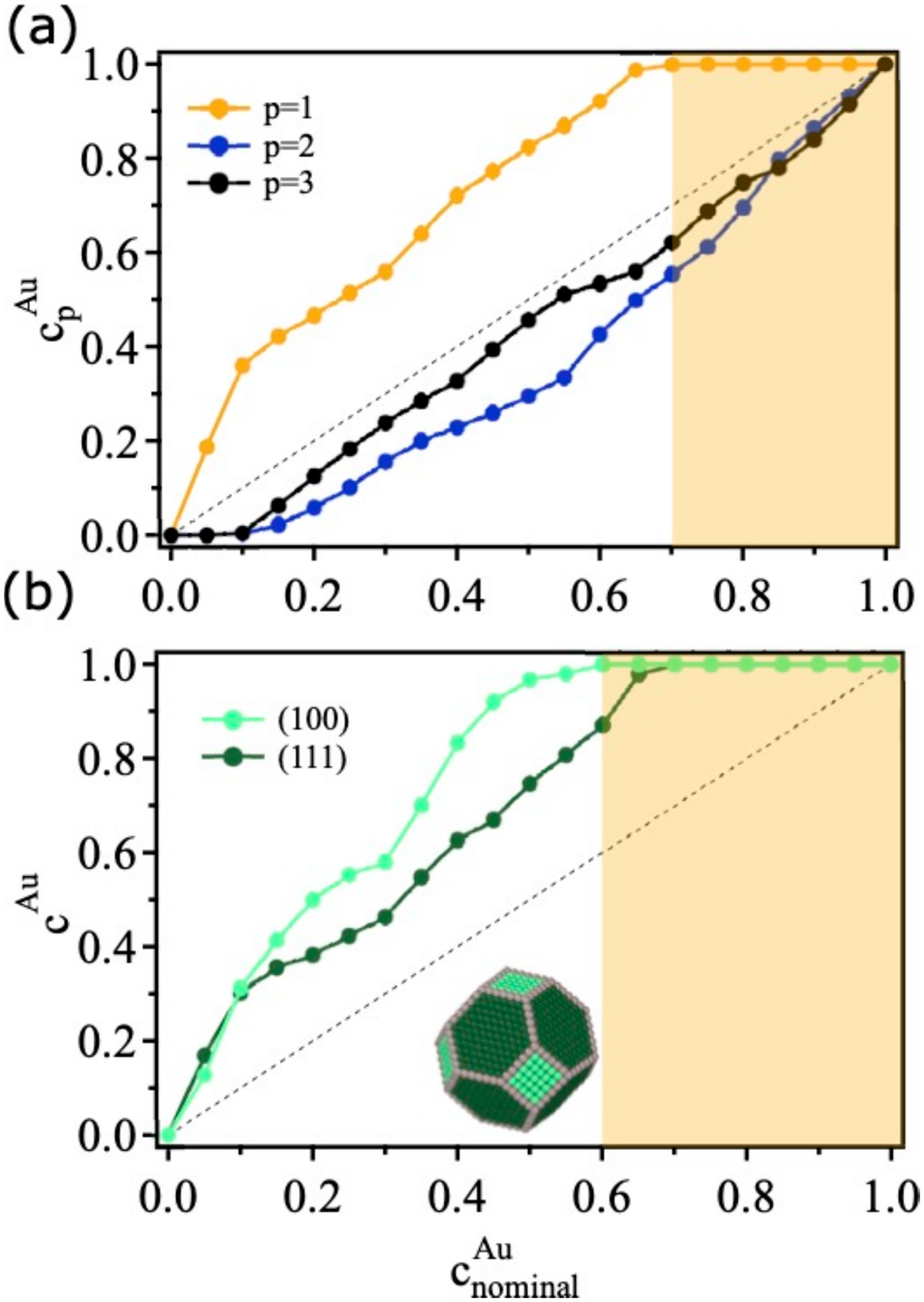}
\caption{Segregation isotherms in the disordered state for (a) the surface ($p=1$) and the first two sublayers ($p=2,3$) and (b) the (100) and (111) surfaces for the truncated octahedron Cu-Au nanoalloys containing 4033 atoms.}
\label{Isotherms}
\end{figure}

We considered truncated octahedron Cu-Au NPs containing 405, 1289 and 4033 atoms. This corresponds to cluster sizes around 2 to 5 nm close to the range explored experimentally. Fig.~\ref{Isotherms}a depicts the segregation isotherms for a Cu-Au NP in a disordered state containing 4033 atoms with initial composition covering the whole phase diagram.
Obviously, our simulations show that Au segregates whatever its nominal concentration with complete segregation when the gold concentration exceeds 70\%. Concomitantly, we observe a depletion in Au atoms in the first two sublayers. Beyond that, a progressive return to the nominal concentration is achieved. To go further, surface concentrations of different facets are analysed in Fig.~\ref{Isotherms}b. Interestingly, there is an enhancement of Au enrichment when going from close-packed (111) facet to the open (100) one in agreement with usual surface energy arguments~\cite{Treglia1999}. Gold enrichment is therefore strongly favored on a (100) surface compared to a (111) one, resulting in a complete saturation from a nominal concentration of 60 at. \% gold.
\begin{figure}[htbp!]
\includegraphics[width=1.00\linewidth]{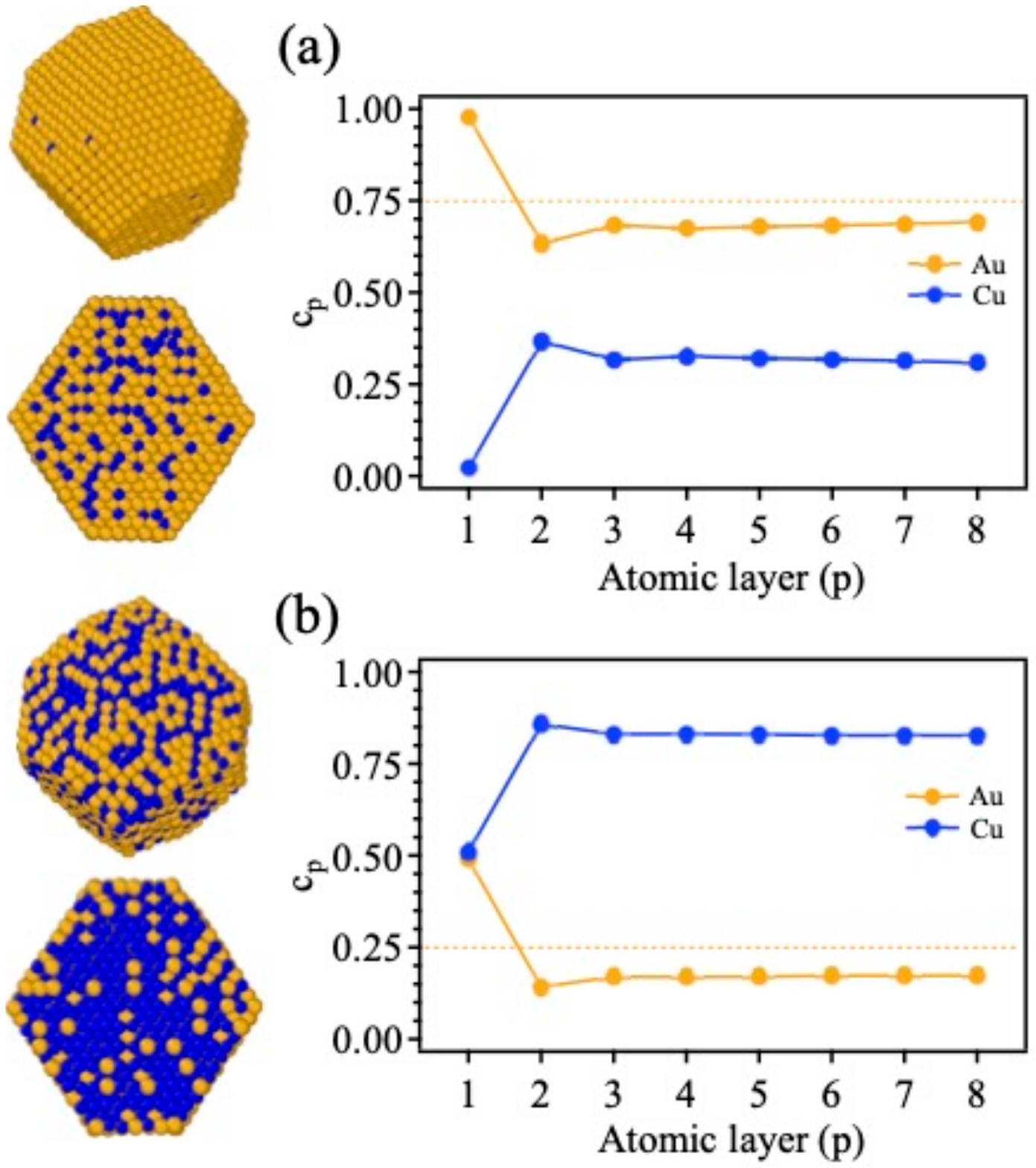}
\caption{Cross-section views of a characteristic equilibrium configuration and concentration profiles along the radius of a NP containing 4033 atoms after performing MC simulations in the disordered state. (a) CuAu$_{3}$ and (b) Cu$_{3}$Au.}
\label{concentration_profile}
\end{figure}

We now focus on specific concentrations where ordered phases exist, i.e. CuAu$_{3}$ and Cu$_{3}$Au~\cite{Okamoto1987}. Let start with the CuAu$_{3}$ composition in a disordered state. After performing MC simulations, strong segregation effects are highlighted. A visual inspection depicts a NP completely surrounded by a thin layer of gold as seen in Fig.~\ref{concentration_profile}a. The innermost layers do not show any particular gold enrichment but display a totally random structure typical of a disordered state. This is confirmed in a more quantitative way with the analysis of the density profiles along the radius of the NPs. The first layer is completely enriched in gold then the concentration decreases within the NP to get closer to the nominal concentration of 75\% of Au. In a second step, Cu$_{3}$Au NPs are addressed and exhibit a much less striking segregation effect. As seen in Fig.~\ref{concentration_profile}b, only 50\% of the surface is covered with Au. Although a significant enrichment of the surface in gold is observed (about twice the nominal concentration), our atomistic simulations do not show a complete Au layer surrounding the NP. Consequently, atomistic simulations confirm that Au segregates at the surface of NPs whatever their composition and size. 

To compare quantitatively the experimental measurements to the numerical calculations, we extracted, from the atomistic simulations with a truncated octahedron containing 4033 atoms, the concentration profiles of each plane from the surface in the same way that they are acquired in the TEM. For that purpose, we determine for the Cu$_{15}$Au$_{85}$ composition, the concentration of each column of the (200) planes. The composition profiles for each atomic column, for different line scans, are superimposed to the experimental curves in Fig.~\ref{Cu15Au85}b. The same procedure was applied for the Cu$_{70}$Au$_{30}$ composition (see Fig.~\ref{Cu70Au30}b). For both compositions, we show a remarkably good quantitative agreement between both results. In particular, due to the intrinsic statistical nature of the FCC disordered phase, the experimental curves are situated in between the two envelope curves determined from the numerical model. Moreover, we prove without any ambiguity the effect of Au segregation on these planes for these alloy stoichiometries. \\


Knowing the real composition of the surface of bimetallic NPs is crucial for their applications. Regarding Cu$_{x}$Au$_{1-x}$ NPs, we succeeded in showing the segregation of a single layer of Au at the surface thus addressing an existing debate in the literature. This has been achieved by combining measurements of the surface composition of (100) and (111) surfaces of NPs with very well controlled morphology and atomic scale simulations. Moreover, our detailed analysis have proven a concentration profile from the facets to the core of the NP as already discussed in case of infinite surfaces highlighting that such mechanism is still present at the nanoscale. These conclusions are very important for the surface and catalyst community~\cite{Cui2013} since it highlights how it is crucial to consider the real surface composition and the concentration profile along the NPs to analyze the reactivity of small catalysts. This work constitutes a major step showing that the future to understand catalysis of nanoalloys is to determine the complete concentration profile of the NPs surfaces under real environmental conditions since segregation behavior can even be changed~\cite{Zhu2013}.


\section*{Supplemental material : \\Atomic Scale Segregation in Copper-Gold Nanoparticles}

\textbf{Sec. I. Synthesis of nanoparticles and preparation for HRTEM characterization}\\

To clearly evidence the effect of Cu or Au segregation, a perfect control on the 3D morphology of the NPs is mandatory. In this context, we developed the epitaxial growth of Cu-Au NPs. The substrate of choice for a good epitaxy of Cu-Au system is NaCl because of its cubic symmetry together with one of its lattice distances (d$_{110}$= 0.398 nm) which is in the same range as the one of Cu-Au alloy (0.385 nm for the Cu$_{50}$Au$_{50}$ composition). Twonominal compositions were made using pulsed laser deposition technique~\cite{Alloyeau2007}: Cu$_{3}$Au and CuAu$_{3}$ and with a nominal thickness of the continuous film fixed at 0.7 nm. The NPs were deposited on a 1 cm$^{2}$ freshly cleaved sodium chloride NaCl (001) single-crystal surface. 
\begin{figure}[htbp!]
\includegraphics[width=0.90\linewidth]{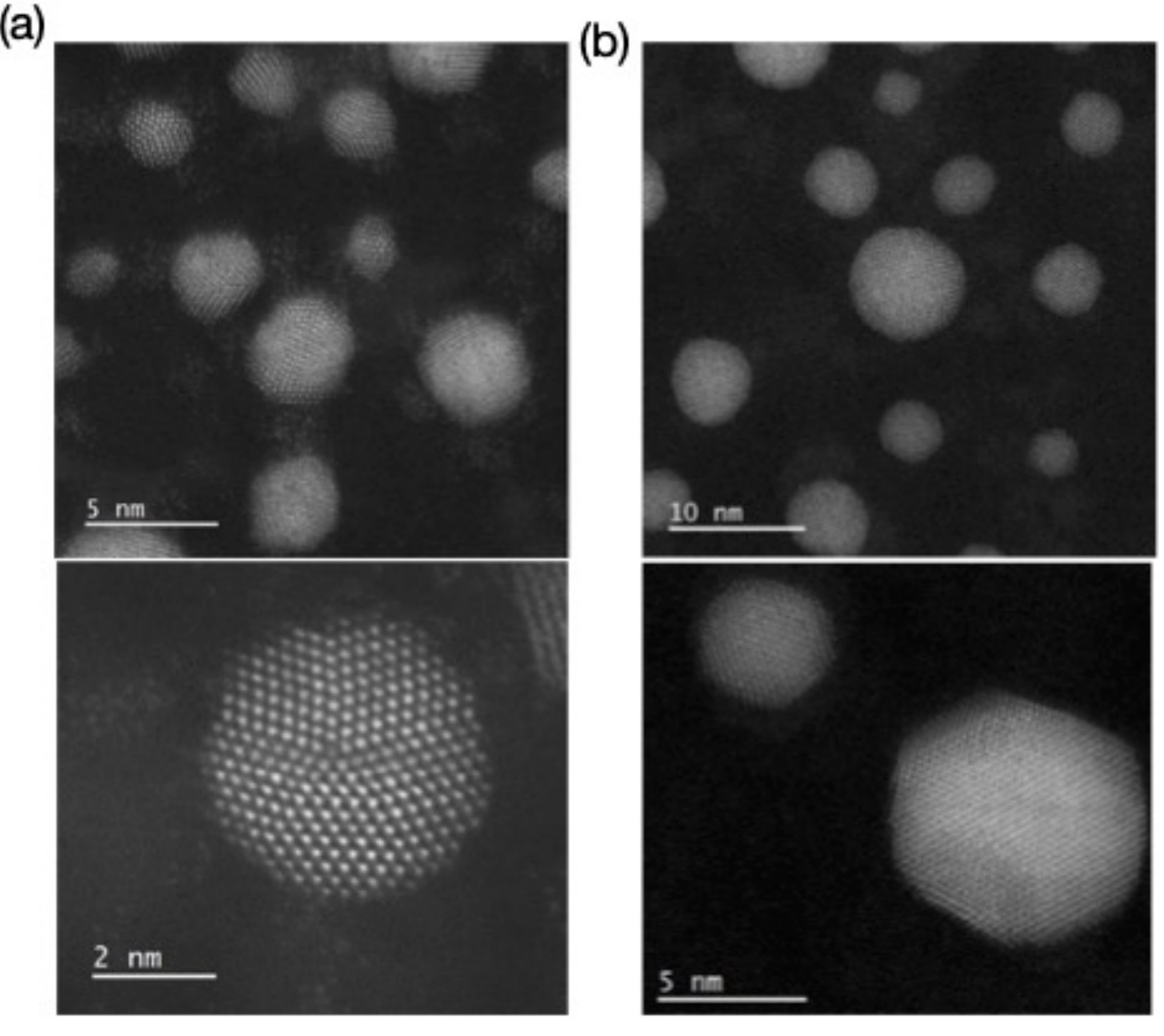}
\caption{HRSTEM images of AuCu$_{3}$ NPs synthesized (a) without epitaxial growth and (b) from epitaxial growth. }
\label{non_epitaxy}
\end{figure}
To ensure well-oriented growth of the NPs on the substrate and in a face-centered cubic (FCC) disordered state (A1 phase), the latter was heated at 400\degree C during the laser deposition. After deposition, the Cu-Au NPs were covered by a few nm-thick amorphous (a-)carbon film obtained by evaporating a carbon rod in an Edward Auto 306 thermal evaporator. Carbon replica~\cite{PierronBohnes2014} was obtained by dissolving the sodium chloride in deionized water followed by the transfer of the carbon-supported NPs to standard Mo grids (300 mesh, Agar Scientific) for structural investigations by TEM. In Fig.~\ref{non_epitaxy}, we can clearly see how epitaxial growth allows us to obtain perfectly facetted and very well oriented NPs limiting the presence of twinning type defects and thus greatly facilitating chemical analysis.   \\

%
%

\textbf{Sec. II. Quantification analysis of the composition of an atomic column}\\

\begin{figure}[htbp!]
\includegraphics[width=0.75\linewidth]{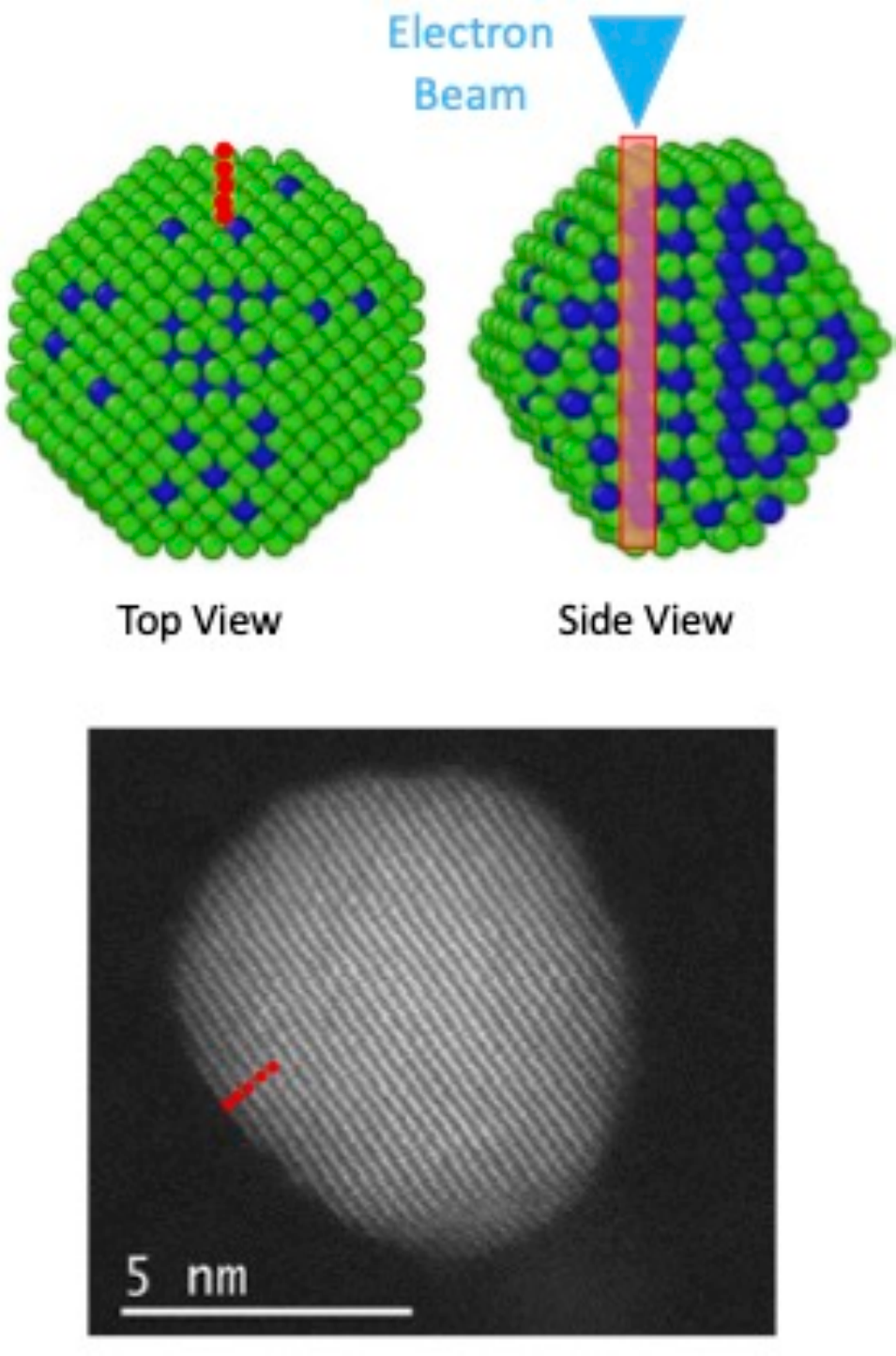}
\caption{Schematic representation of the EDX measurement.}
\label{EDX}
\end{figure}
As seen in Fig.~\ref{EDX}, EDX measurement consists of a succession of acquisition points along a line. The distance between two points is chosen so that it coincides with the distance between two planes of the considered family ($\sim$0.2 nm). Each acquisition lasts 20 seconds and irradiates the whole column below the acquisition point, before moving on to the next one, i.e. the next plane. A drift corrector correcting the position every 5 seconds is also in place due to the magnification ($\times$10M) of the image.\\

\begin{figure}[htbp!]
\includegraphics[width=1.0\linewidth]{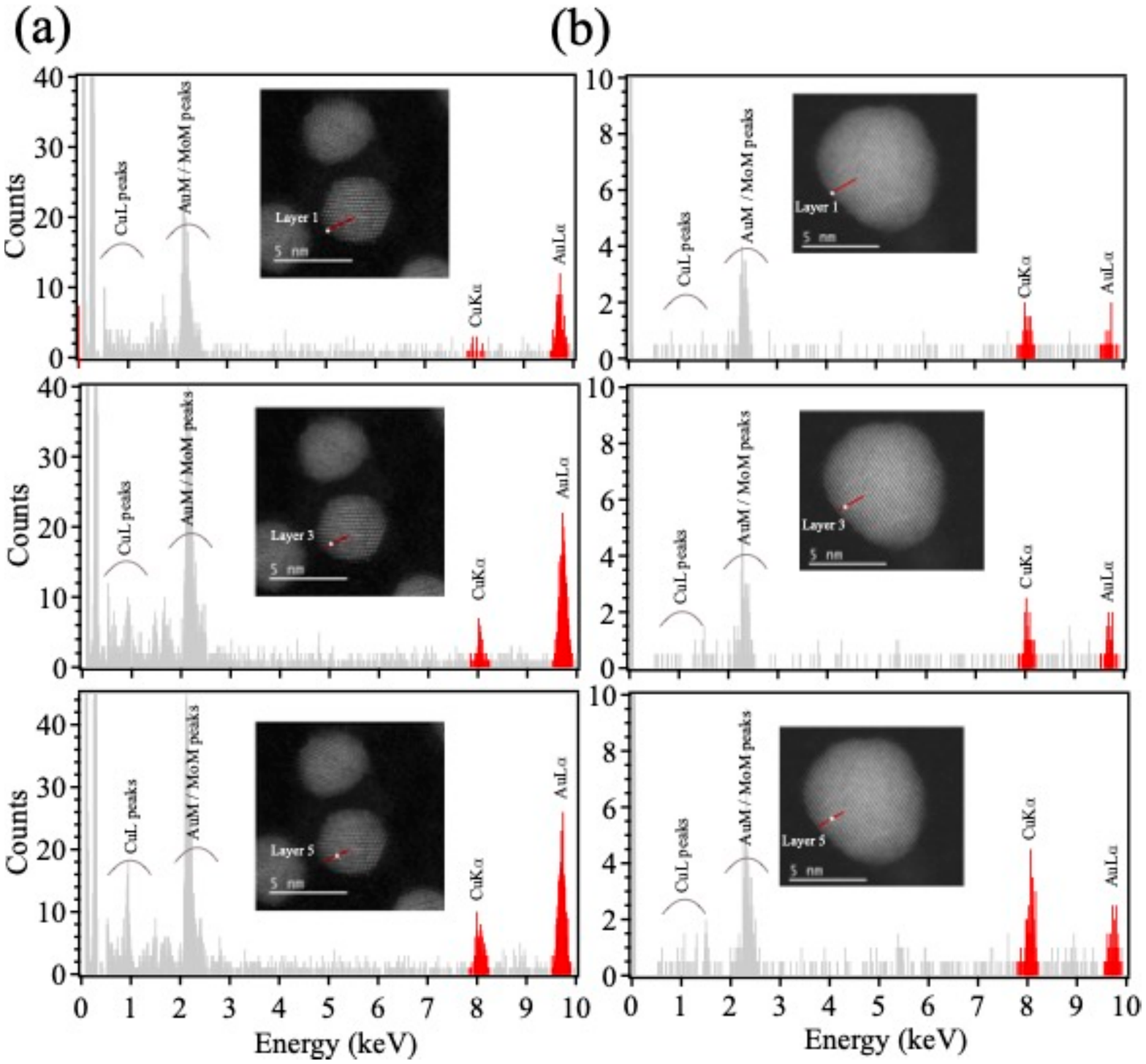}
\caption{EDX spectra acquired on atomic columns belonging to the 1$^{st}$, 3$^{rd}$ and 5$^{th}$ atomic planes (from top to bottom) from the surface along the line perpendicular to the NP surface (red line on the image of the inset): (a) Cu$_{15}$Au$_{85}$ and (b) Cu$_{70}$Au$_{30}$.}
\label{EDX_Scan}
\end{figure}

In Fig.~\ref{EDX_Scan}, we present typical EDX spectra corresponding to the position of the electron beam on an atomic column of planes parallel to the facet along the line of acquisition. The composition of the columns (hence the one of the planes) was obtained by quantifying the intensity of the Cu K$_{\alpha}$ and Au L$_{\alpha}$ peaks. We do not used the Au M peaks appearing at lower energies since there is an overlap between those peaks and the Mo M ones coming from the support of the carbon thin film supporting the NPs. The background level in these spectra is very low because the NPs are deposited on a very thin carbon film. The noise in this kind of data follows a Poisson statistic and thus the error bar in Fig. 1b and 2b was calculated, for each plane along the line of analysis, using the standard error propagation approach. Finally, the configuration parameters of the microscope for the acquisition of HRSTEM image and EDX spectra were fix as follow: condenser aperture of 40 $\mu$m, spot size 8C and camera length of 5 cm, so that the resolution (i.e. the probe size) can be roughly estimated to be less than 0.1 nm.
\\
\\
\\
\\

\textbf{Sec. III. TB-SMA interatomic potential} \\

$\bullet$ \textbf{TB-SMA model} \\

Within the tight-binding (TB) framework~\cite{ducastelle1970, rosato1989}, the total energy of an atom $n$ is splitted in two parts, a band structure term that describes the formation of an energy band $E_{band}^{n}$ when atoms are put together and a repulsive term $E_{rep}^{n}$ that empirically accounts for the ionic and electronic repulsions : $E_{tot}^{n} = E_{band}^{n} + E_{rep}^{n} $. The total energy of the system containing $N$ atoms, $E_{tot}$, then writes :
\begin{equation}
E_{tot}=\displaystyle\sum_{n=1,N} E_{tot}^{n}.
\end{equation}
In the SMA formalism, the band energy is given by :
\begin{equation}
E_{band}^{n} = -\sqrt{\sum_{m\neq n}\xi^{2}_{ij}\exp\left[-2q_{ij}\left(\frac{r_{nm}}{r^{0}_{ij}}-1\right)\right]}
\end{equation}
and the repulsive contribution is described in a pairwise Born-Mayer form :
\begin{equation}
E_{rep}^{n} = \sum_{m\neq n} A_{ij} \exp\left[-p_{ij}\left(\frac{r_{nm}}{r^{0}_{ij}}-1\right)\right]
\end{equation}
where $r_{nm}$ is the distance between atoms at sites $n$ and $m$ whereas $r_{ii}^{0}$ (respectively $r_{jj}^{0}$) corresponds to the equilibrium distance between first neighbors in the pure metal $i$ (respectively $j$), $r_{ij}^{0}=\frac{r_{ii}^{0}+r_{jj}^{0}}{2}$ corresponds to the equilibrium distance between first neighbors in the alloy. Note that $\xi_{ij}$ is the effective hopping integral between atoms $i$ and $j$. \\

$\bullet$ \textbf{The fitting procedure} \\

In the present work, the parameters ($\xi_{ij}$, $A_{ij}$, $q_{ij}$ and $p_{ij}$) are fitted to reproduce several bulk physical properties. 
\begin{table}[!ht]
\centering
\begin{tabular}{c|c|c|c|c}
\hline
\hline
   	   & $\xi_{ij}$  & $A_{ij}$ 	&$p_{ij}$	&$q_{ij}$ \\
   	   & (eV) & (eV)	&	& \\ \hline
Cu-Cu &1.29&0.09&11.06&2.46	\\ 
Au-Au  &1.80& 0.21	&10.95	&4.02 \\ 
Cu-Au &1.61 & 0.14	&10.68	&3.24\\ \hline
 \hline
\end{tabular}
\caption{Parameters of the interatomic potentials used for the Cu-Cu, Au-Au and Cu-Au interactions.}
\label{parameters}
\end{table}
The resulting parameter values for Cu-Cu, Au-Au and Cu-Au interactions are presented in Table~\ref{parameters}. For pure elements, the TB-SMA parameters have been fitted on experimental values for the fcc structure namely the lattice parameter, the cohesive energy and the elastic moduli (bulk modulus and the two shear moduli) as seen in Table~\ref{fitpur}. \\

\begin{table}[!htbp]
\centering
\begin{tabular}{c|c|c|c|c|c|c|c|c}
\hline
\hline
   	   & Lattice   & Cohesive   &B	&C'&C$_{44}$ & $\gamma_{111}$ & $\gamma_{100}$ & $\gamma_{110}$\\
   	   & parameter & energy 	& 	&  & & & & \\ \hline
Cu &3.62  &-3.50 &142 & 24 	&75 &1.10 & 1.18 & 1.32 \\ 
    & (3.62) & (-3.50)& (142)&  (26)	& (75) & (1.95) &  (2.17)& (2.24) \\ 
Au  &4.08 & -3.81 	&166  &16 &45 & 0.51   & 0.59 & 0.63 \\ 
      &(4.08)& (-3.81)	& (166) & (16)& (45)&  (1.28)  &  (1.63) & (1.70) \\ 
\hline
 \hline
\end{tabular}
\caption{For pure Au and Cu elements (fcc structure), comparison of our SMA model with experimental data (lattice parameters (\AA)~\cite{Kittel1995}, cohesive energies (eV/at)~\cite{Kittel1995}, elastic modulii (GPa)~\cite{Simmons1971}, and surface energies (J.m$^{2}$)~\cite{Vitos1998}) indicated in bracket.}
\label{fitpur}
\end{table}
Note that fitting the TB-SMA parameters to experimental cohesive energies leads to an underestimation ($\sim$ by a factor 2) of the surface energies since it is well known that such potentials are not always adapted to describe physical properties from bulk to surface~\cite{brooks1983}. Nevertheless, these deviations do not prevent us from describing qualitatively structural properties of Cu-Au nanoalloys as we will see in the following. Improving the accuracy usually implies increasing the number of parameters, which can blur the physical transparency of the model~\cite{guevara1995, barreteau2002, goyhenex2012}. Typically, the following hierarchy: $\gamma_{111}<\gamma_{100}<\gamma_{110}$ is well reproduced within our TB-SMA model for both Cu and Au surfaces. Moreover, we can notice that the Au surface energy calculated is lower than the one of Cu, again in agreement with DFT calculations, which is in favor of the Au surface segregation. \\

For the Cu-Au interaction, the potential has been fitted to the enthalpies of solution in the two diluted limits. Our results (-0.19 eV and -0.12 eV) are in good agreement with experimental data~\cite{Hultgren1973} (-0.21 eV and -0.10 eV) in case of Cu(Au) and Au(Cu), respectively. Moreover, we check that the TB-SMA model can reproduce the enthalpy of segregation of the solute at the surface, $\Delta H^{seg}$ defined as:
\begin{equation}
\Delta H^{seg} = E^{tot}_{surface}(solute)-E^{tot}_{bulk}(solute)
\label{eqn:DH_seg}
\end{equation}
where $E^{tot}_{surface}(solute)$ (respectively, $E^{tot}_{bulk}(solute)$) is the total energy of the system when one solute atom is at a surface site (respectively, a bulk site). A negative value of $\Delta H^{seg}$ means that solute segregation is favored at the surface. To get a relevant database  for checking the validity of our TB-SMA model, we have performed DFT calculations to determine $\Delta H^{seg}$.  The projector-augmented wave (PAW) method~\cite{Blochl1994, Kresse1999}, as implemented in the Vienna \textit{ab initio} simulation package (VASP) code~\cite{Kresse1993} was used. The generalized-gradient approximation functional of the exchange correlation energy was calculated within the Perdew, Burke, and Ernzerhof formulation (GGA-PBE)~\cite{Perdew1992}. The cut-off energy was fixed at 400 eV and the positions of the atoms in the supercell are relaxed until the total electronic energy differences fall below 10$^{-6}$ eV. Au and Cu surfaces were calculated using slab models with a $3\times3$ surface unit cell, five layers and a vacuum region of 15 \AA. The two-bottom layer were frozen in the relaxed bulk positions, while the topmost four layers were allowed to relax. The Brillouin-zone integrations for surfaces were performed on a Monkhorst-Pack ($3\times3\times1$) \textit{k}-point mesh.

\begin{table}[!htbp]
\centering
\begin{tabular}{c|c|c|c}
\hline
\hline
& (100)   & (111)  & (110) \\
\hline
Au atom in Cu matrix & -0.23  & -0.22 &  -0.25\\
 &  (-0.37) & (-0.32)  & (-0.43) \\
Cu atom in Au matrix &  0.15  & 0.12 & 0.08 \\
 & (0.46) &  (0.42) & (0.37)\\
\hline
 \hline
\end{tabular}
\caption{Segregation energies (in eV) at different surface sites of one Au (Cu) solute in a Cu (Au) matrix from our TB-SMA potential. DFT values are indicated in bracket.}
\label{segregation}
\end{table}
As seen in Table.~\ref{segregation}, although our TB-SMA model is not quantitatively perfect, it still succeeds in reproducing the significant trends revealed by the DFT calculations. In the case of Au (Cu) solute, our calculations show a negative (positive) value of $\Delta H^{seg}$ for the first layer whatever the surface considered. This is in agreement with the DFT data suggesting that gold segregation is favored for all surfaces unlike Cu which has the opposite behavior.\\

%
%


$\bullet$ \textbf{Monte Carlo simulations} \\

This atomic interaction model is  implemented in a Monte Carlo (MC) code in the canonical ensemble, based on the Metropolis algorithm, which allows to relax the structures at finite temperature~\cite{Frenkel2002}. In the canonical ensemble, MC trials correspond to random displacements of randomly chosen atoms and exchanges between two randomly chosen atoms of different species. The average quantities are calculated over $10^{6}$ MC macrosteps, a similar number of macrosteps being used to reach equilibrium.  A MC macrostep corresponds to $N$ propositions of chemical switches and $N$ propositions of random atomic displacements, $N$ being the total number of atoms of the cluster. \\

%
%
\begin{acknowledgments}

H.A. thanks B. Legrand for fruitful discussions. 

\end{acknowledgments}

\end{document}